\documentclass[11pt,a4paper]{article}
\pdfoutput=1

\usepackage{jcappub}
\usepackage{amsmath,amssymb,color,bm, graphicx}
\usepackage{enumerate,xstring, hyperref}

\def\be{\begin{equation}}
\def\ee{\end{equation}}
\def\bea{\begin{eqnarray}}
\def\eea{\end{eqnarray}}

\def\ba#1\ea{\begin{align}#1\end{align}}
\def\bg#1\eg{\begin{gather}#1\end{gather}}

\newcommand{\refeq}[1]{Eq.~(\ref{eq:#1})}

\newcommand{\refsec}[1]{Sec.~\ref{sec:#1}}

\def\Pnl{P_\text{m}}

\renewcommand{\v}[1]{\bm{#1}}

\newcommand{\vx}{\v{x}}

\newcommand{\vk}{\v{k}}

\newcommand{\vtheta}{\v{\theta}}
\newcommand{\vell}{\v{\ell}}

\def\be{\begin{equation}}
\def\ee{\end{equation}}
\def\ben{\begin{eqnarray}}
\def\een{\end{eqnarray}}
\def\ba{\begin{array}}
\def\ea{\end{array}}

\def\ba#1\ea{\begin{align}#1\end{align}}

\newcommand{\bq}{\begin{eqnarray}}
\newcommand{\eq}{\end{eqnarray}}
\newcommand{\bes}{\begin{subequations}}
\newcommand{\ees}{\end{subequations}}

\def\P{\mathcal{P}}
\def\W{\mathcal{W}}

\makeatletter
\newlength{\apb@width}
\newcommand{\autoparbox}[2][c]{\settowidth{\apb@width}{#2}\parbox[#1]{\apb@width}{#2}}

\makeatother

\DeclareMathOperator{\cov}{Cov}
\DeclareMathOperator{\bfcov}{\bf Cov}
\DeclareMathOperator{\covkappa}{Cov_{\kappa}}

\newcommand{\comment}[1]{}

\begin{document}

\title{Accurate cosmic shear errors: do we need ensembles of simulations?}

\author{Alexandre Barreira,$^1$}
\emailAdd{barreira@MPA-Garching.MPG.DE}
\affiliation{$^1$Max-Planck-Institut f{\"u}r Astrophysik, Karl-Schwarzschild-Str.~1, 85741 Garching, Germany}

\author{Elisabeth Krause,$^{2,3}$ and}
\emailAdd{krausee@email.arizona.edu}
\affiliation{$^2$Steward Observatory, University of Arizona, 933 N Cherry Ave, Tucson, AZ 85721, U.S.A.\newline
$^3$California Institute of Technology, 1200 E California Blvd, Pasadena, CA 91125, U.S.A.}

\author{Fabian Schmidt$^1$}
\emailAdd{fabians@MPA-Garching.MPG.DE}

\abstract{Accurate inference of cosmology from weak lensing shear requires an accurate shear power spectrum covariance matrix. Here, we investigate this accuracy requirement and quantify the relative importance of the Gaussian (G), super-sample covariance (SSC) and connected non-Gaussian (cNG) contributions to the covariance. Specifically, we forecast cosmological parameter constraints for future wide-field surveys and study how different covariance matrix components affect parameter bounds. Our main result is that the cNG term represents only a small and potentially negligible contribution to statistical parameter errors: the errors obtained using the G+SSC subset are within $\lesssim 5\%$ of those obtained with the full G+SSC+cNG matrix for a Euclid-like survey. This result also holds for the shear two-point correlation function, variations in survey specifications and for different analytical prescriptions of the cNG term. The cNG term is that which is often tackled using numerically expensive ensembles of survey realizations. Our results suggest however that the accuracy of analytical or approximate numerical methods to compute the cNG term is likely to be sufficient for cosmic shear inference from the next generation of surveys.}


\date{\today}

\maketitle
\flushbottom


\section{Introduction}\label{sec:intro}

Ongoing (e.g.~KiDS \cite{2017MNRAS.465.1454H, 2017arXiv170706627J, 2017arXiv170605004V}, DES \cite{diehl/etal:2014, 2017arXiv170801530D}, HSC \cite{2018PASJ...70S..25M}) and future (e.g.~Euclid \cite{2011arXiv1110.3193L}, LSST \cite{2012arXiv1211.0310L}, WFIRST\cite{2013arXiv1305.5422S}) large imaging surveys have been and are expected to keep setting ever tighter constraints on various competing cosmological models. The comparison between theory and observations requires a likelihood function $\mathcal{L}({\bf D}|{\bf M}({\bf S}))$ to quantify the probability that the observed data vector ${\bf D}$ is a statistical realization of some cosmological model with parameters ${\bf S}$ and associated data vector prediction ${\bf M}(\bf S)$. Under the common assumption that the data vector is Gaussian distributed, we can write:

\bq\label{eq:L}
\mathcal{L}({\bf D}|{\bf M}({\bf S})) = \frac{1}{\sqrt{(2\pi)^d{\rm det}(\bfcov)}}{\rm exp}\left[-\frac{1}{2}\left({\bf M}({\bf S}) - {\bf D}\right)^{t}\bfcov^{-1}\left({\bf M}(\bf S) - {\bf D}\right)\right], \nonumber \\
\eq
where $d$ is the size of the data vector ${\bf D}$. The posterior probability distribution of cosmological parameters is given via Bayes' theorem as $\P({\bf S}|{\bf D})  \propto \mathcal{L}({\bf D}|{\bf M}({\bf S}))\P({\bf S})$, where $\P(\bf S)$ is some prior probability function on the parameters. In addition to the observed data vector and theoretical prediction, parameter inference and goodness-of-fit analyses also require the covariance matrix ${\bfcov}$, which at the end of the day is what controls the size of the error bars on parameters. In this paper, we discuss the accuracy requirements for the covariance matrix of two-point weak lensing statistics (cf.~Eq.~(\ref{eq:estimator}) below), or in other words, how well do we need to know the covariance in order to meet a desired uncertainty on the uncertainty of estimated parameters.

The covariance matrix of lensing two-point statistics can be organized into three physically distinct types of contributions which we refer to as the Gaussian (G), super-sample covariance (SSC) and connected non-Gaussian (cNG) terms (we describe these terms more carefully in Sec.~\ref{sec:covdec}). The G term is the minimal covariance contribution (technically, the disconnected part of the four point function) and it would be the only contribution to the covariance if the noisy shear field itself was Gaussian distributed; this is approximately correct on sufficiently large scales (multipoles $\ell \lesssim 100-200$ for galaxy source redshifts $z_S \approx 1$). The SSC term \cite{takada/hu:2013, li/hu/takada, 2014PhRvD..90j3530L, 2014PhRvD..90b3003M, 2014MNRAS.441.2456T, 2014MNRAS.444.3473T} describes the correlation between the two-point function on different scales that is induced by large scale density/tidal fluctuations in which the entire surveyed region is embedded in. Finally, the cNG term \cite{1999ApJ...527....1S, 2001ApJ...554...56C, 2009MNRAS.395.2065T, 2016JCAP...06..052B, 2016PhRvD..93l3505B, mohammed1, responses1, responses2} describes the contribution to the covariance that is induced by nonlinear structure formation within the survey volume, i.e., when the density fluctuations grow to order unity and the field becomes appreciably non-Gaussian distributed. The Gaussian term can be calculated given the survey footprint and the nonlinear matter power spectrum, which cosmic emulators \cite{emulator} can now predict to close to $1 \%$ precision for a range of cosmological parameter values. The SSC term can also be fully specified by the survey footprint and the power spectrum, as well as the so-called first-order power spectrum responses to density and tidal fields \cite{completessc}, which can be efficiently measured with separate universe simulations \cite{li/hu/takada, 2014PhRvD..90j3530L, wagner/etal:2014, CFCpaper2, li/hu/takada:2016, lazeyras/etal, response, andreas}. The cNG term is controlled by the so-called parallelogram configuration of the nonlinear matter trispectrum \cite{1999ApJ...527....1S} (the Fourier transform of the  matter four-point correlation function), which is difficult to evaluate; the cNG term is thus the least well understood part of the covariance.

There are two main ways to tackle the calculation of the cNG term: analytical approaches and the so-called ensemble approach. Analytical approaches target the direct evaluation of the matter trispectrum. The simplest such example relies on standard or effective perturbation theory \cite{1999ApJ...527....1S, 2016JCAP...06..052B, bertolini1, 2016PhRvD..93l3505B}, but this comes with the drawback of being predictive only on fairly large scales. Approaches based on the halo model \cite{cooray/sheth} are another popular way to evaluate the cNG term: the total matter trispectrum is split into 1-, 2-, 3- and 4-halo contributions \cite{cooray/hu:2001_cov,2009MNRAS.395.2065T}. Under the model assumptions, this yields a calculation that is valid deep in the nonlinear regime of structure formation; the drawback, however, is that the assumptions of the halo model are known to result only in a rough approximation to the power spectrum, and presumably, to the trispectrum as well (see e.g.~Ref.~\cite{2016PhRvD..93f3512S} for a discussion). Recently, Ref.~\cite{responses2} used the so-called response approach to perturbation theory described in Ref.~\cite{responses1} to calculate the parallelogram trispectrum. The response approach consists of an extension of perturbation theory that is valid in the nonlinear regime of structure formation for the case of squeezed interaction vertices, i.e.~it describes the coupling of large-scale quasi-linear modes to two small-scale fully nonlinear ones. The work of Ref.~\cite{responses2} illustrated that the trispectrum is dominated by such squeezed interactions, and in particular, that the response approach can be used to capture the bulk (roughly $70$\%) of the total trispectrum.

The nontriviality of predicting the trispectrum accurately from first principles has motivated the implementation of an alternative and computationally highly intensive approach based on ensembles. The main idea in this approach is to generate a sufficiently large number of statistically independent realizations of the matter density field (or projected weak lensing shear field), and then measure the power spectrum sample covariance across this ensemble \cite{2009ApJ...700..479T, 2009ApJ...701..945S, 2011ApJ...734...76S, 2012MNRAS.426.1262H, li/hu/takada, blot2015, 2017ApJ...850...24T, 2016PhRvD..93f3524P, 2018MNRAS.478.4602K, 2018arXiv180504511H}. Each realization of the nonlinear three-dimensional density field requires performing one N-body simulation, which makes these methods computationally very demanding. Recently, some efforts have been devoted to developing approximate and fast N-body based methods to alleviate the computational burden \cite{2016MNRAS.459.2327I, 2018MNRAS.478.4602K, 2018arXiv180105745S, 2017JCAP...01..008R}, but these methods come with the price of reduced accuracy. The case is less severe for realizations of the cosmic shear field since outputs of the same (large-volume) N-body simulation can be recycled to produce several quasi-independent lensing maps; for instance, Ref.~\cite{2016PhRvD..93f3524P} shows that a single N-body simulation may be sufficient to generate $\sim 10^4$ such realizations of the shear field. The analysis of future surveys will, nonetheless, require ensemble sizes that are a few orders of magnitude larger than what has been done to date in order to estimate sufficiently accurate covariance matrices (see e.g.~Ref.~\cite{2007A&A...464..399H, 2013MNRAS.432.1928T} for a discussion). Hence, although not prohibitive, the calculation of lensing covariances with ensembles remains a computationally intensive task under the requirements of future surveys. We note also that analytical approaches are effectively noise-free.

The main result of this paper is that the cNG term represents only a small contribution to the final errors on cosmological parameters inferred from two-point statistics of cosmic shear. We demonstrate this for the case of a cosmic-shear-only analysis with survey parameters similar to the specifications expected for Euclid (case $s01$ in Table \ref{table:setups} below) and for LSST (case $s02$), as well as for a number of variations around these configurations. We also always work under the assumption of Gaussian distributed data (we comment further on this in Sec.~\ref{sec:conc}). We show in particular that, for all the cases tested, the cNG term is only responsible for an increase in the size of the statistical error bars of $< 8\%$ (this is the precision on the error bars, not on the parameters). In the analysis presented in this paper, we always neglect any systematic uncertainties that one would marginalize over in real-world analyses, and which would further increase the errors on parameters. This means the estimates of the importance of the lensing cNG term presented here are rather conservative. Our results thus suggest that the accuracy of analytical or approximate N-body based cNG calculations is likely sufficient for the analysis of future cosmic shear data.

This paper is organized as follows. In Sec.~\ref{sec:method}, we describe the lensing convergence tomographic data vector and the corresponding expressions of the G, SSC and cNG covariance terms. Section \ref{sec:results} displays our main results for the impact of using various covariance decompositions on the parameter constraints. We summarize and discuss the consequences of our findings in Sec.~\ref{sec:conc}.

\section{Weak lensing data vector and covariance matrix}\label{sec:method}

In this section, we describe the survey specifications and data vectors that we consider in our forecast analyses, as well as the lensing covariance matrix contributions that we wish to test. 

\subsection{Lensing data vector}\label{sec:datav}

Throughout this paper, we take as observables two-dimensional maps of the weak lensing convergence field estimated from source galaxies in some redshift bin $i$ (see e.g.~Refs.~\cite{2001PhR...340..291B, 2005astro.ph..9252S, 2008ARNPS..58...99H, 2015RPPh...78h6901K} for lensing reviews):
\bq
\label{eq:kappa1} \kappa^i_\W(\vtheta) &=& \W(\vtheta) \kappa^i(\vtheta), \\
\label{eq:kappa2} \kappa^i(\vtheta) &=& \int_0^\infty {\rm d}\chi g^i(\chi) \delta(\vx = \chi \vtheta, z(\chi)),
\eq
where $\delta$ denotes the total matter density contrast, $\vtheta$ is the angular coordinate on the sky, $z(\chi)$ is the redshift at comoving distance $\chi$ and the lensing kernel is given by 
\bq\label{eq:lenskernel}
g^i(\chi) = \frac{3H_0^2\Omega_{\rm m}}{2c^2} (1+z(\chi)) \chi \int_\chi^\infty {\rm d}\chi' \frac{n_S^i(z(\chi'))}{\bar{n}_\mathrm{eff}^i}\frac{{\rm d}z}{{\rm d}\chi'} \frac{\chi'-\chi}{\chi'},
\eq
where $n_S^i(z)$ is the redshift distribution of lensing source galaxies in bin $i$, $\bar{n}^i_\mathrm{eff}$ the projected effective source galaxy density in bin $i$, $\Omega_{\rm m}$ the present-day fractional total matter density and $H_0$ the present-day Hubble expansion rate; we also always assume a spatially-flat Friedmann-Robertson-Walker spacetime. We label by $N_{\rm tomo}$ the number of source galaxy redshift bins. In Eq.~(\ref{eq:kappa1}), $\W(\vtheta)$ represents the survey window function (or mask/footprint; we use these words interchangeably), which in this paper we consider to be contiguous, unity inside the surveyed area and zero outside. 

As a data vector, we consider the angle-averaged auto- and cross-power spectra of the $N_{\rm tomo}$ lensing maps (tomographic power spectra), which can be obtained with the following estimator
\bq\label{eq:estimator}
\hat{C}^{ij}_\kappa(\ell_1) = \frac{1}{\Omega_\W}\int_{\Omega_{\ell_1}}\frac{{\rm d}^2\vell}{2\pi\ell_1\Delta_{\ell_1}}\tilde{\kappa}^i_\W(\vell)\tilde{\kappa}^j_\W(-\vell),
\eq
where $\tilde{\kappa}^i_\W(\vell)$ are lensing convergence Fourier amplitudes measured from the maps (a tilde indicates a Fourier space quantity), the integration range $\Omega_{\ell_1}$ is an annulus with width $\Delta_{\ell_1}$ centered at $\ell_1$ and $\Omega_\W = 4\pi f_{\rm sky} = \int {\rm d}^2\vtheta \W(\vtheta)$ is the survey area; $f_{\rm sky}$ is the surveyed total sky fraction. The theoretical prediction for the expectation value of the tomographic convergence power spectrum can be obtained via
\bq\label{eq:prediction}
C_\kappa^{ij}(\ell) = \int_0^{\infty} {\rm d}\chi \frac{g^i(\chi)g^j(\chi)}{\chi^2} \Pnl(k_{\ell}, z(\chi)),
\eq
where $k_{\ell} = (\ell + 1/2)/\chi$. We evaluate the nonlinear three-dimensional matter power spectrum $\Pnl$ using the revised {\sc Halofit} \cite{2003MNRAS.341.1311S} fitting formula of Ref.~\cite{2012ApJ...761..152T}. Equation (\ref{eq:prediction}) assumes the flat-sky and Limber's approximations, which are sufficient for the multipoles $\ell \gtrsim 20$ considered in this analysis.

For the redshift distribution of the source galaxies $n_S(z)$ we consider both Euclid- and LSST-like distributions. For Euclid, we follow Ref.~\cite{2013LRR....16....6A} and take $n_S(z) = z^2{\rm exp}\left[-\left(z/z_0\right)^{3/2}\right]$ ($z_0 = z_{\rm mean}/1.412$ and $z_{\rm mean} = 0.9$) with a projected effective source density of $\bar{n}_\mathrm{eff} = 30\ {\rm arcmin}^{-2}$. The LSST source distribution is based on the simulations of Ref.~\cite{chang/etal:2013}, including updates described in the LSST Dark Energy Science Collaboration Science Requirement Document \cite{DESC-SRD}, which yields $\bar{n}_\mathrm{eff} = 26\ {\rm arcmin}^{-2}$. These distributions are split into $N_{\rm tomo}$ tomographic bins each with the same number of galaxies; for $N_{\rm tomo} = 3,5,10$ the data vector corresponds, respectively, to 6, 15 and 55 auto/cross-spectra. In each of these, we label by $N_{\ell}$ the number of $\ell$ bins between some minimum and maximum values $\ell_{\rm min}$ and $\ell_{\rm max}$, respectively, equally spaced in log-scale. 

In our results below, we consider both noise-free data vectors matching Eq.~(\ref{eq:prediction}) evaluated at the fiducial cosmology, as well as noisy realizations of the data vector with Eq.~(\ref{eq:prediction}) as the mean and with multivariate Gaussian noise drawn from one of the covariance matrices.

For completeness, we note that cosmic shear analyses are frequently carried out using two-point shear correlation functions in configuration space. Here, we opt to perform our analysis with the lensing convergence $\kappa$ and in Fourier space for simplicity of calculation and because it suffices to illustrate our main conclusions. The corresponding shear correlation function predictions can be obtained from the lensing convergence via additional integrations over Bessel functions
\bq\label{eq:xis}
\xi^{ij}_{+/-}(\theta) = \int_0^{\infty} \frac{{\rm d}\ell}{2\pi} \ell J_{0/4}(\ell\theta) C_\kappa^{ij}(\ell),
\eq
where $\xi^{ij}_{+/-}(\theta)$ are the two shear cross-correlation functions between tomographic bins $i$ and $j$. Thus, the $\xi^{ij}_{+/-}(\theta)$ are effectively linear combinations of the $C_\kappa^{ij}(\ell)$. The conclusions we draw in this paper for $C_{\kappa}^{ij}(\ell)$ data vectors with $\ell \in \left[\ell_{\rm min}, \ell_{\rm max}\right]$ will thus hold for $\xi^{ij}_{+/-}(\theta)$ on the range of angular scales $\theta$ that are accurately described by angular wavenumbers $\ell \in \left[\ell_{\rm min}, \ell_{\rm max}\right]$. We will explicitly verify that our conclusions on the unimportance of the cNG term from the power spectrum analysis hold also in the correlation function case.

\subsection{Lensing covariance decomposition}\label{sec:covdec}

In lensing covariance related work (see e.g.~Refs.~\cite{takada/hu:2013, completessc, 2018arXiv180511629T} for a few recent examples), it has become customary to decompose the total covariance matrix of the estimator of Eq.~(\ref{eq:estimator}) into three terms\footnote{This decomposition is only strictly valid if we assume the Limber approximation for the sub-survey modes involved (see Appendix D of Ref.~\cite{completessc} for a discussion). The Limber approximation is nonetheless valid for the modes $\ell \gtrsim 20$ we consider in this paper.} known as the Gaussian (G), connected non-Gaussian (cNG) and the super-sample covariance (SSC, which is also of connected and non-Gaussian nature, but is restricted to the effect of super-survey modes):
\bq\label{eq:covdecom}
\covkappa^{ijmn}(\ell_1, \ell_2) &=& \Big<\hat{C}^{ij}_\kappa(\ell_1)\hat{C}^{mn}_\kappa(\ell_2)\Big> - \Big<\hat{C}^{ij}_\kappa(\ell_1)\Big>\Big<\hat{C}^{mn}_\kappa(\ell_2)\Big> \\
&=& \covkappa_G^{ijmn}(\ell_1, \ell_2) + \covkappa^{ijmn}_{cNG}(\ell_1, \ell_2) + \covkappa^{ijmn}_{SSC}(\ell_1, \ell_2).
\eq
Next, we briefly summarize the main equations associated with these three terms; we adopt the same notation as in Ref.~\cite{completessc}, to which (and references therein) we refer the reader for more details about the calculation of the lensing covariance that we use in this paper.

\subsubsection{The G term}\label{sec:Gterm}

The G covariance term is given by
\bq\label{eq:Gterm}
\covkappa_G^{ijmn}(\ell_1, \ell_2) &=& \frac{4\pi\delta_{\ell_1\ell_2}}{\Omega_\W(2\ell_1+1)\Delta\ell_1} \left[\Bigg(C_\kappa^{im}(\ell_1) + \delta_{im}\frac{\sigma_e^2}{2\bar{n}^i_{\mathrm{eff}}}\Bigg)\Bigg(C_\kappa^{jn}(\ell_1) + \delta_{jn}\frac{\sigma_e^2}{2\bar{n}^j_{\mathrm{eff}}}\Bigg)  \right. \nonumber \\
&&\left. \ \ \ \ \ \ \ \ \ \ \ \ \ \ \ \ \ \ \ \ +  \Bigg(C_\kappa^{in}(\ell_1) + \delta_{in}\frac{\sigma_e^2}{2\bar{n}^i_{\mathrm{eff}}}\Bigg)\Bigg(C_\kappa^{jm}(\ell_1) + \delta_{jm}\frac{\sigma_e^2}{2\bar{n}^j_{\mathrm{eff}}}\right) \Bigg],\quad
\eq
where $\sigma_e = 0.37$ is the RMS ellipticity of the source galaxies ($\bar{n}_{\mathrm{eff}}^i = \bar{n}_{\mathrm{eff}}/N_\mathrm{tomo}$ for all tomographic bins). The Kronecker deltas $\delta_{\ell_1\ell_2}$ and $\delta_{im}$ ensure the Gaussian term is non-vanishing only if $\ell_1$ and $\ell_2$ are in the same $\ell$ bin and the shape noise terms only contribute for matching galaxy tomographic bins. For simplicity, we do not consider the effect of the mask shape on shape noise \cite{2011ApJ...734...76S, 2018arXiv180410663T}.

Equation (\ref{eq:Gterm}) receives its name because it is the only contribution that arises if the density field is Gaussian distributed; at later stages in structure formation this is not the case, but the functional form of Eq.~(\ref{eq:Gterm}) remains the same (with the power spectrum being the nonlinear one). 
Overall, given its straightforward dependence on the theoretical prediction, the Gaussian term is well understood and not a source of big concern in covariance studies of angular power spectra. The survey footprint does not change the fundamental ingredients in \refeq{Gterm}, although the covariance becomes non-diagonal \cite{2004MNRAS.349..603E, 2002ApJ...567....2H}. These (contiguous) mask convolution effects can however be ignored for the sufficiently small angular scales (relative to the typical angular size of the footprint) and multipole binning we consider in our analysis.

\subsubsection{The cNG term}\label{sec:cNGterm}

The cNG term describes the correlations between different sub-survey modes that exist if the density field is non-Gaussian distributed \cite{1999ApJ...527....1S}, which is the case at late times during nonlinear structure formation or at all times in cosmologies with primordial non-Gaussianity. Under the Limber and flat-sky approximations (again, valid on $\ell \gtrsim 20$), this term can be written as
\bq\label{eq:cNGterm}
\covkappa_{cNG}^{ijmn}(\ell_1, \ell_2) &=& \frac{1}{\Omega_\W} \int_0^{2\pi}\frac{{\rm d}\varphi_{\vell_1}}{2\pi} \int_0^{2\pi}\frac{{\rm d}\varphi_{\vell_2}}{2\pi} \int_0^{\infty} {\rm d}\chi \frac{g^i(\chi)g^j(\chi)g^m(\chi)g^n(\chi)}{\chi^6} \nonumber \\ 
&& \times T_{\rm m}(\vk_{\ell_1}, -\vk_{\ell_1}, \vk_{\ell_2}, -\vk_{\ell_2}; z(\chi)),
\eq
where $T_{\rm m}$ is the equal-time matter trispectrum (the subscript $_{\rm m}$ in $T_{\rm m}$ should not be confused with the tomographic bin superscript $^m$), i.e., the Fourier transform of the equal-time connected four-point matter correlation function:
\bq\label{eq:trispectrum}
\langle\delta(\vk_a)\delta(\vk_b)\delta(\vk_c)\delta(\vk_d)\rangle_c = (2\pi)^3 T_{\rm m}(\vk_a, \vk_b, \vk_c, \vk_d) \delta_D(\vk_a + \vk_b + \vk_c + \vk_d).
\eq
In Eq.~(\ref{eq:cNGterm}), $\varphi_{\vell}$ denotes the polar angle of the vector $\vell$ and we have trivially performed the bin averages by assuming the trispectrum does not vary rapidly within each bin.

In the results below, we adopt two different recipes to evaluate the matter trispectrum. The first, which we consider as the default recipe (cf.~Table \ref{table:setups}), is that presented in Ref.~\cite{responses2} based on the response approach to perturbation theory described in Ref.~\cite{responses1}. More specifically, we include the totality of the resummed trispectrum at tree-level, as well as the resumation of the dominant terms at 1-loop level. We do not repeat the formulae here, but the interested reader can find the final expressions in Eqs.~(2.10), (3.3), (3.4) and (4.5) of Ref.~\cite{responses2}; this is the same calculation of the cNG term used in Ref.~\cite{completessc}. In squeezed configurations, e.g., $k_{\ell_1} \ll k_{\ell_2}$ and with $\vk_{\ell_1}$ in the linear regime, the response prediction is guaranteed to capture the totality of the matter trispectrum up to corrections that scale as $(k_{\ell_1}/k_{\ell_2})^2$. When both modes are in the nonlinear regime, the response prediction captures $\approx 70\%$ of $T_{\rm m}$ in the parallelogram configuration as measured in Ref.~\cite{blot2015} using an ensemble of over 12000 N-body simulations; as explained in more detail in Ref.~\cite{responses2}, the accuracy of the response approach can be improved in regimes when both $\vk_{\ell_1}$,$\vk_{\ell_2}$ are in the nonlinear regime by including the rest of the 1-loop term with perturbation theory and including also 2-loop terms.

To guard against biased conclusions based on inaccuracies of the response calculation of the cNG term (in the regime where both modes are nonlinear), we also compute a covariance matrix using the halo model formalism \cite{cooray/sheth} (cf.~$s08$ case in Table \ref{table:setups}). More specifically, we use the recipe presented in Ref.~\cite{2009MNRAS.395.2065T} for the concrete application of lensing covariance matrices, on which the cNG terms used in the real data analyses of the KiDS \cite{2017MNRAS.465.1454H} and DES \cite{2017arXiv170609359K, 2017arXiv170801530D} surveys are also based. The details of our halo model implementation and modeling choices for the halo model ingredients are described in Ref.~\cite{2017MNRAS.470.2100K}. The halo model has known deficiencies that follow directly from the simplifying model assumptions (see e.g.~Ref.~\cite{2016PhRvD..93f3512S} for a recent discussion), and as a result, it is also not guaranteed to be an accurate description of the true trispectrum. It is however a physically motivated framework whose regime of validity is not the same as that of the response based calculation, and which we can thus use to cross-check that our conclusions on the relative size of the cNG term are not peculiar to the response calculation. 

Before proceeding, we also point out the ensemble approach to the cNG term \cite{2009ApJ...700..479T, 2009ApJ...701..945S, 2011ApJ...734...76S, 2012MNRAS.426.1262H, li/hu/takada, blot2015, 2017ApJ...850...24T, 2018arXiv180504511H}, which, as discussed in \refsec{intro}, proceeds by estimating the cNG term using a large ensemble of statistically independent realizations of the power spectrum\footnote{With the ensemble approach one also measures the Gaussian contribution as well, but not the SSC term if the boxes where the power spectrum is measured are not embedded in larger-scale fluctuations.}. Here, we do not consider cNG contributions estimated in this way, but a take-away point of the results in the next section is that the cNG term has limited impact on parameter inference, which might obviate the need for expensive ensembles dedicated to measuring it.

\subsubsection{The SSC term}\label{sec:SSCterm}

The super-sample covariance term\footnote{This term  encompasses the so-called {\it beat coupling} \cite{2006MNRAS.371.1188H, 2006PhRvD..74b3522S, 2009MNRAS.395.2065T, 2012JCAP...04..019D} and {\it halo sample variance} \cite{2007NJPh....9..446T, 2013MNRAS.429..344K, 2009ApJ...701..945S, 2003ApJ...584..702H} contributions discussed in previous literature.} describes the coupling between observed sub-survey and unobserved super-survey modes, i.e., modes whose wavelengths are larger than the surveyed volume. For flat-sky lensing applications in the Limber approximation, the derivation of this term follows straightforwardly from matter trispectrum terms that get excited by the finite size of the survey window function \cite{takada/hu:2013, completessc}. Reference~\cite{completessc} presents a derivation of the lensing SSC term that goes beyond the Limber approximation in the super-survey modes (for which the approximation can be insufficient) and that is valid for general curved-sky masks (see also Ref.~\cite{2016arXiv161205958L}). For the case of the angle-averaged tomographic lensing convergence power spectrum, this term is given by
\bq
\label{eq:SSCterm1}\covkappa_{SSC}^{ijmn}(\ell_1, \ell_2) &=& \frac{1}{\Omega_\W^2} \sum_{LM} |b_{LM}|^2 \sigma_{\ell_1, \ell_2}^{L, ijmn} \\
\label{eq:SSCterm2}\sigma_{\ell_1, \ell_2}^{L, ijmn} &=& \frac{2}{\pi} \int_0^{\infty} {\rm d}p p^2 P_\text{L}(p, z=0) f_{\ell_1}^{L, ij}(p)f_{\ell_2}^{L,mn}(p),
\eq
where $P_\text{L}$ is the linear matter power spectrum and 
\bq\label{eq:flLp}
f_{\ell}^{L, ij}(p) = \int_0^{\infty} {\rm d}\chi \frac{g^i(\chi)g^j(\chi)}{\chi^2} D(z)\Pnl(k_{\ell}, z) \Big(R_1(k_{\ell}, z) + \frac{R_K(k_{\ell}, z)}{6} + \frac{1}{2}R_K(k_{\ell}, z)\partial^2_x\Big)j_L(x) ,\nonumber \\
\eq
where $D(z)$ is the linear growth factor (we assumed $z \equiv z(\chi)$ to ease the notation; also the subscript $_\mathrm{L}$ in $P_\text{L}$ should not be confused with the mask angular wavenumber $L$) and $x = p\chi$. In Eq.~(\ref{eq:SSCterm1}), the $b_{LM}$ are the spherical harmonic coefficients of the survey mask (defined on the curved sky; we use {\sc Healpix}\footnote{http://healpix.sf.net} to evaluate mask power spectra). The functions $R_1(k_{\ell}, z)$ and $R_K(k_{\ell}, z)$ denote, respectively, the first-order power spectrum responses to long-wavelength isotropic density and tidal field  perturbations. The isotropic response has been measured with separate universe simulations in Ref.~\cite{li/hu/takada:2016} (and subsequently also in Refs.~\cite{wagner/etal:2014, response}); the tidal response has been measured only more recently in Ref.~\cite{andreas} using a generalization of the separate universe technique to anisotropic cosmologies (see also Ref.~\cite{2018JCAP...02..022L} for measurements of the tidal response for galaxy mocks using sub-volumes of a larger volume simulation).

\section{Results}\label{sec:results}

\begin{table}
  \caption{Summary of the survey specifications considered in this paper. The case labeled as $s01$ corresponds to the expected specifications for a Euclid-like survey; the case $s02$ represents a LSST-like survey with $N_{\rm tomo} = 5$. The cases $s03-s13$ correspond to variations around the $s01$ setup ($s03-s12$ all have $N_{\rm tomo} = 5$, $s13$ has $N_{\rm tomo} = 3$). A star indicates that the value is the same as in the main setup $s01$. The units of  $\bar{n}_\mathrm{eff}$ are  ${\rm arcmin}^{-2}$.}
\begin{tabular}{@{}lccccccccccc}
\hline\hline
\\
Setup  & $f_{sky}$ &  Mask & $N_{\rm tomo}$ & $N_{\ell}$ & $\ell_{\rm min}$ & $\ell_{\rm max}$ & cNG & $\bar{n}_\mathrm{eff}$
\\
\hline
\\
$s01$ (Euclid-like)&\ \  $0.36$ &  Polar cap & $10$ & $20$ & $20$ & $5000$ & Responses & 30 \\ 
\\
$s02$ (LSST-like)&\ \  $0.44$ &  $*$ & $5$ & $*$ & $*$ & $*$ & $*$ & $26$ \\
\\
$s03$&\ \  $*$ &  $*$ & $5$ & $*$ & $*$ & $*$ & $*$ & $*$\\
$s04$&\ \  $*$ &  Equatorial band & $5$ & $*$ & $*$ & $*$ & $*$ & $*$ \\
$s05$&\ \  $*$ &  Two polar caps & $5$ & $*$ & $*$ & $*$ & $*$ & $*$ \\
$s06$&\ \  $0.05$ &  $*$ & $5$ & $*$ & $*$ & $*$ & $*$ & $*$ \\
$s07$&\ \  $0.50$ &  $*$ & $5$ & $*$ & $*$ & $*$ & $*$ & $*$ \\
$s08$&\ \  $*$ &  $*$ & $5$ & $*$ & $*$ & $*$ & Halo Model & $*$ \\
$s09$&\ \  $*$ &  $*$ & $5$ & $10$ & $*$ & $*$ & $*$ & $*$ \\
$s10$&\ \  $*$ &  $*$ & $5$ & $30$ & $*$ & $*$ & $*$ & $*$ \\
$s11$&\ \  $*$ &  $*$ & $5$ & $*$ & $500$ & $10000$ & $*$ & $*$ \\
$s12$&\ \  $*$ &  $*$ & $5$ & $*$ & $500$ & $10000$ & Halo Model & $*$ \\
$s13$&\ \  $*$ &  $*$ & $5$ & $*$ & $*$ & $1000$ & $*$ & $*$ \\
\\
$s14$&\ \  $*$ &  $*$ & $3$ & $*$ & $*$ & $*$ & $*$ & $*$ \\
\\
\hline
\hline
\end{tabular}
\label{table:setups}
\end{table}

In this section, we present our main results on the relative importance of the G, cNG and SSC terms at the level of parameter constraints. Our main results correspond to the survey/analysis specifications labeled as $s01$ in Table \ref{table:setups} for $N_{\rm tomo} = 10$ tomographic bins; these results are discussed below in subsection \ref{sec:res1}. We also verify the robustness of our conclusions against variations to these specifications; these are the $s02 - s13$ cases in Table \ref{table:setups} and the results are discussed in subsection \ref{sec:res2}.

We explore a 5 dimensional cosmological parameter space consisting of the present-day total matter density $\Omega_{\rm m}$, the RMS of the linearly extrapolated matter fluctuations $\sigma_8$ at $z=0$ smoothed on scales of $8\ {\rm Mpc}/h$, a two-parameter parametrization of the time evolution of the equation of state of dark energy $w(a) = w_0 + w_a(1-a)$ (with $a = 1/(1+z)$ the scale factor), and the present-day dimensionless Hubble expansion rate $h$ ($H_0 = 100h\ {\rm km/s/Mpc}$). Our adopted fiducial values for these parameters are
\be
\Big\{\Omega_{\rm m}, \sigma_8, w_0, w_a, h \Big\} = \Big\{0.3, 0.8315, -1 , 0 ,0.7 \Big\}\,.
\ee
All results are obtained with the {\sc CosmoLike} package \cite{2017MNRAS.470.2100K} and we assume a multivariate Gaussian shape of the likelihood function (cf.~Eq.~(\ref{eq:L}); we discuss this assumption in Sec.~\ref{sec:conc}) with flat priors on the parameters.

\subsection{The unimportance of the cNG term for future wide-field surveys}\label{sec:res1}

\begin{figure}
        \centering

        \includegraphics[width=\textwidth]{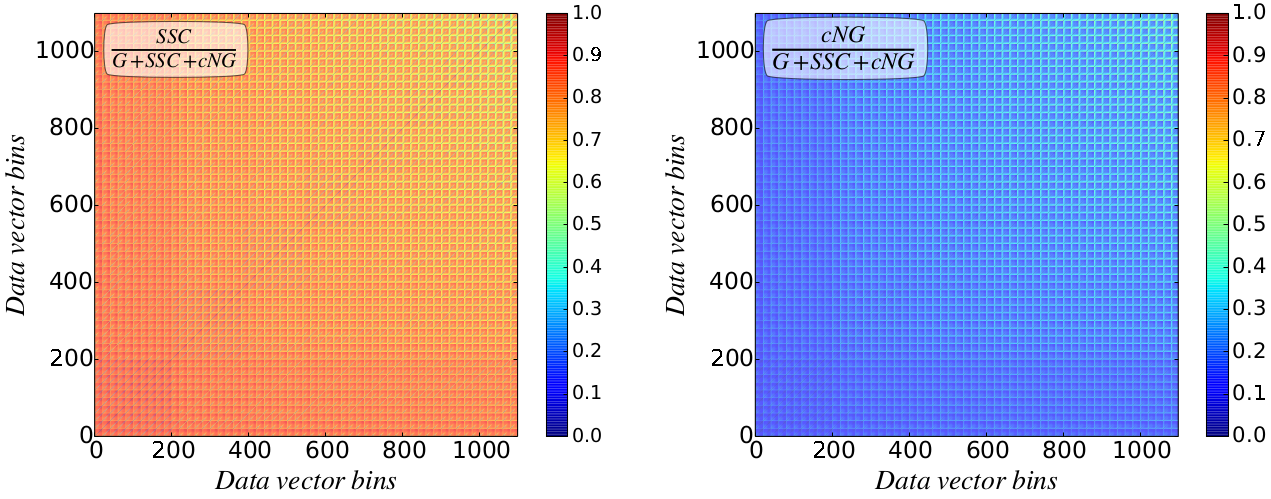}

        \includegraphics[width=\textwidth]{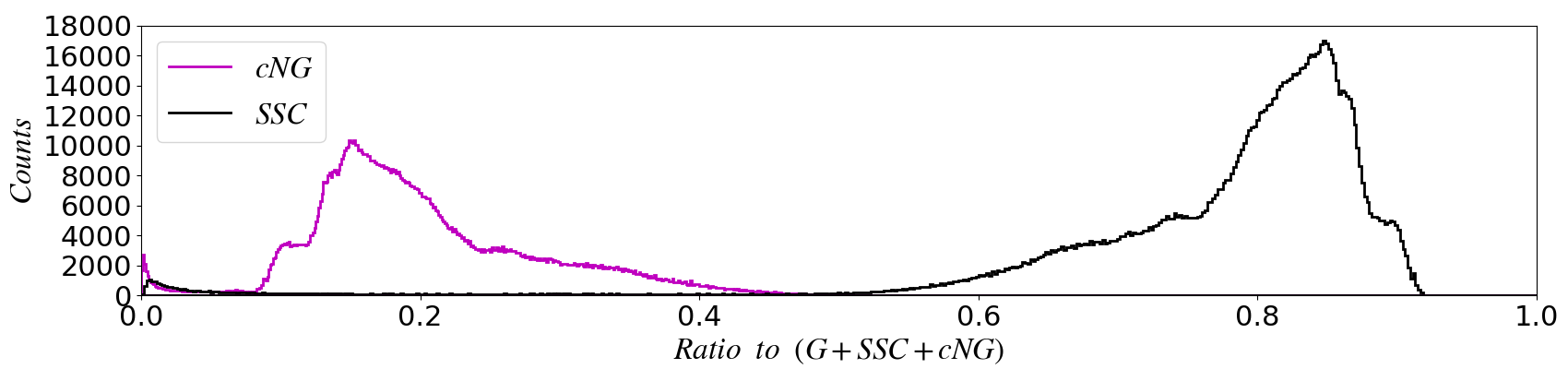}

        \caption{Covariance matrix of the tomographic lensing convergence data vector of our main Euclid-like setup ($s01$ in Table \ref{table:setups}). This corresponds to $N_{\rm tomo} = 10$ and $N_{\ell} = 20$, and hence, the covariance is a $1100\times 1100$ matrix. Each $20\times20$ sub-block contains the covariance of $\ell$ bins for a pair of shear tomography bins, $\cov^{ijmn}(\ell_1, \ell_2)$. The tomography bin indices $(ij)$ increase from left to right/bottom to top, with the ordering $(00),\ldots,(09),(11),\ldots,(19),(22),\ldots,(99)$. Specifically, the upper panels show the element-by-element ratio of the SSC (left) and cNG (right) contributions to the total G+cNG+SSC covariance. The lower panel shows the distribution of the elements of the two matrices, as labeled.}
\label{fig:s01}
\end{figure}

\begin{figure}
        \centering
        \includegraphics[width=\textwidth]{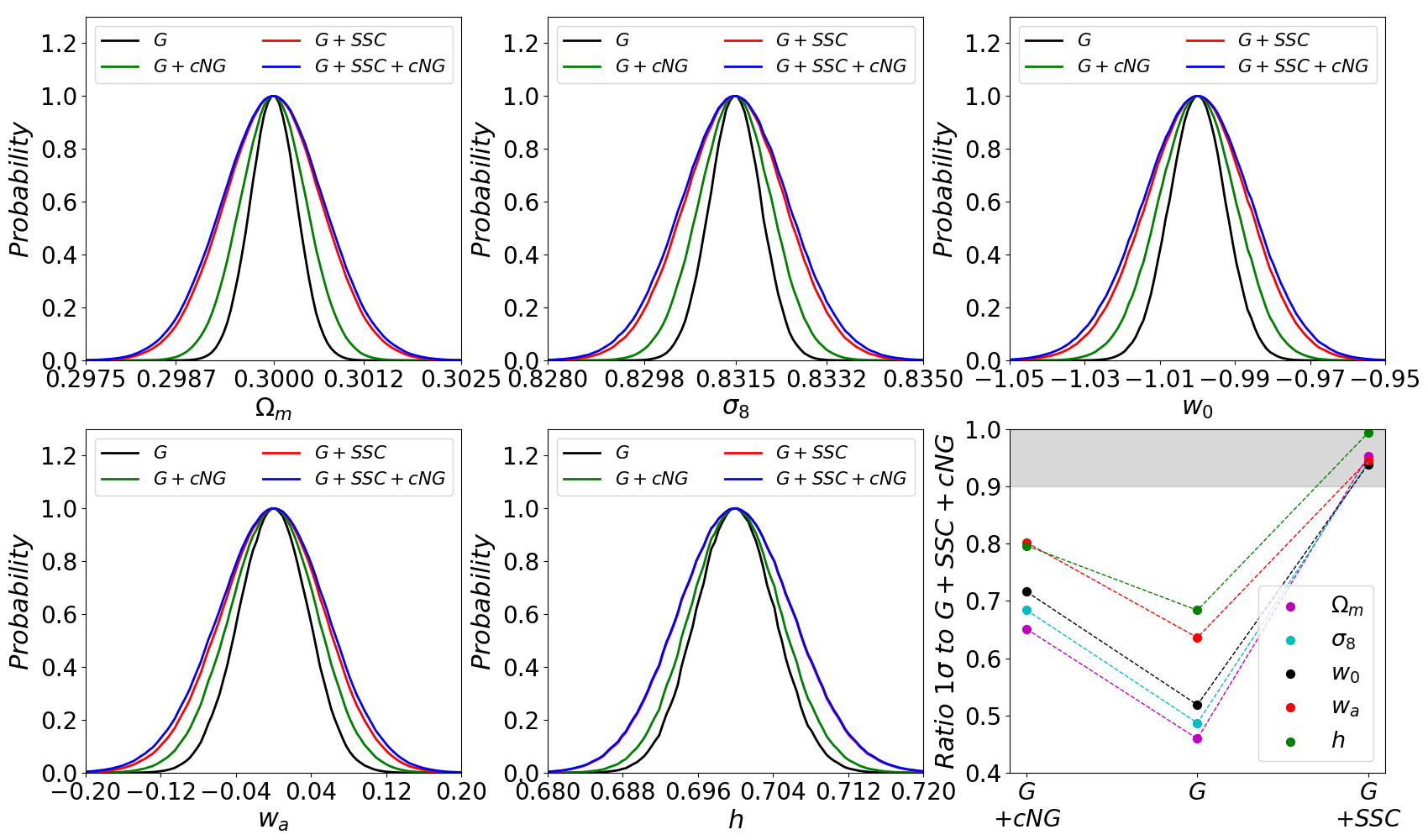}
        \caption{Constraints on cosmological parameters obtained with the main Euclid-like setup ($s01$ in Table \ref{table:setups}), for varying covariance matrix subsets, as labeled. The unmarginalized posteriors shown here for each parameter correspond to constraints obtained with all other parameters held fixed at their fiducial values. The lower right panel shows the ratio of unmarginalized $1\sigma$ confidence intervals obtained with the different covariance subsets to that of the total covariance matrix G+SSC+cNG (the grey band marks $10\%$). The data vector used here corresponds to a noise-free realization matching the prediction of our fiducial cosmology.}
\label{fig:n01_1d}
\end{figure}

\begin{figure}
        \centering
        \includegraphics[width=\textwidth]{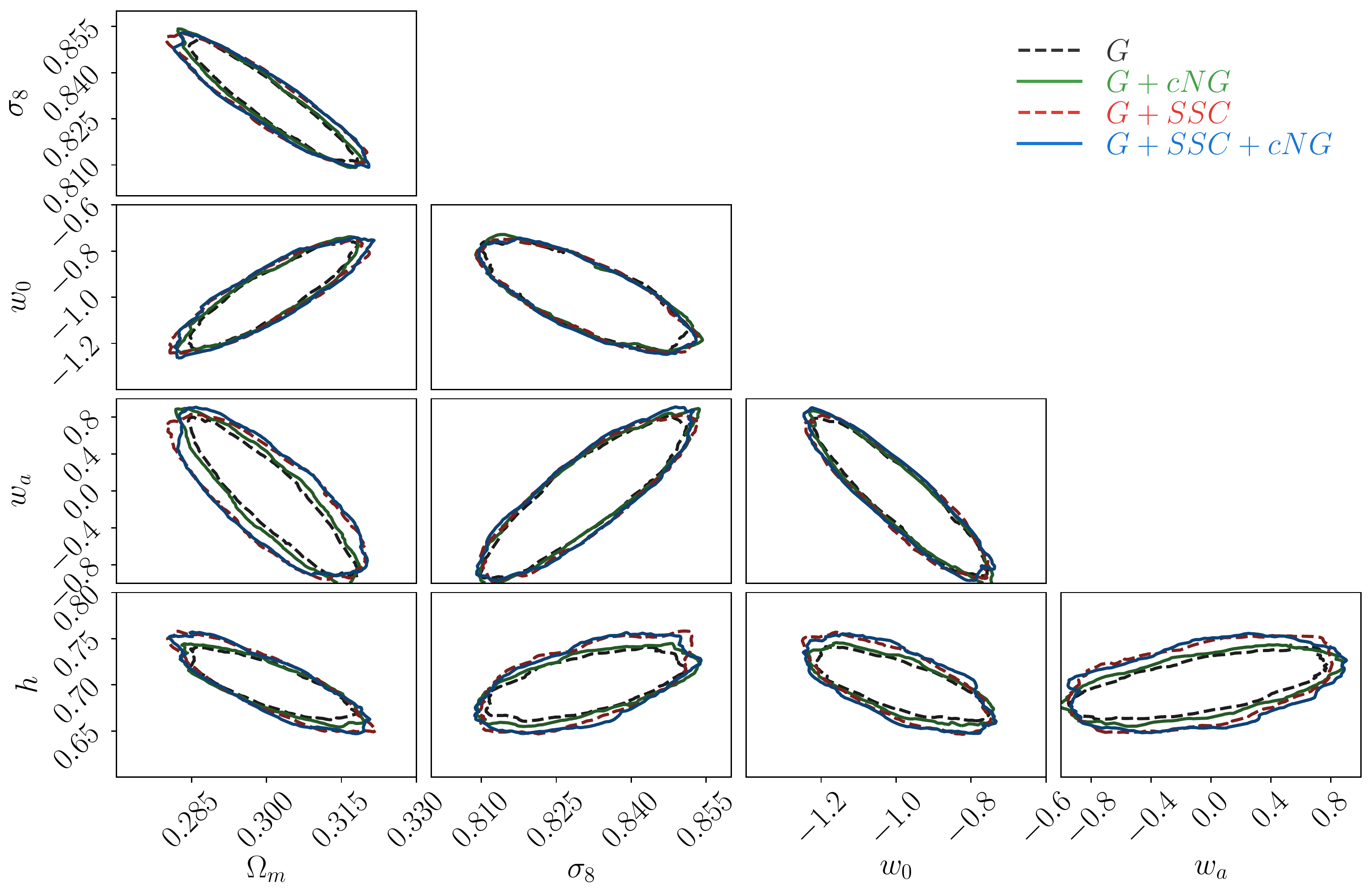}
        \caption{Marginalized two-dimensional $2\sigma$ confidence intervals on cosmological parameters obtained with the main Euclid-like setup ($s01$ in Table \ref{table:setups}), for varying covariance matrix subsets, as labeled. These results were obtained with {\sc Multinest} \cite{2009MNRAS.398.1601F} sampling with all five parameters varying and the resulting chains were processed with the {\rm ChainConsumer} package (https://samreay.github.io/ChainConsumer/index.html). The data vector used here corresponds to a noise-free realization matching the prediction of our fiducial cosmology.
        }
\label{fig:n01_2d}
\end{figure}

\begin{figure}
        \centering
        \includegraphics[width=\textwidth]{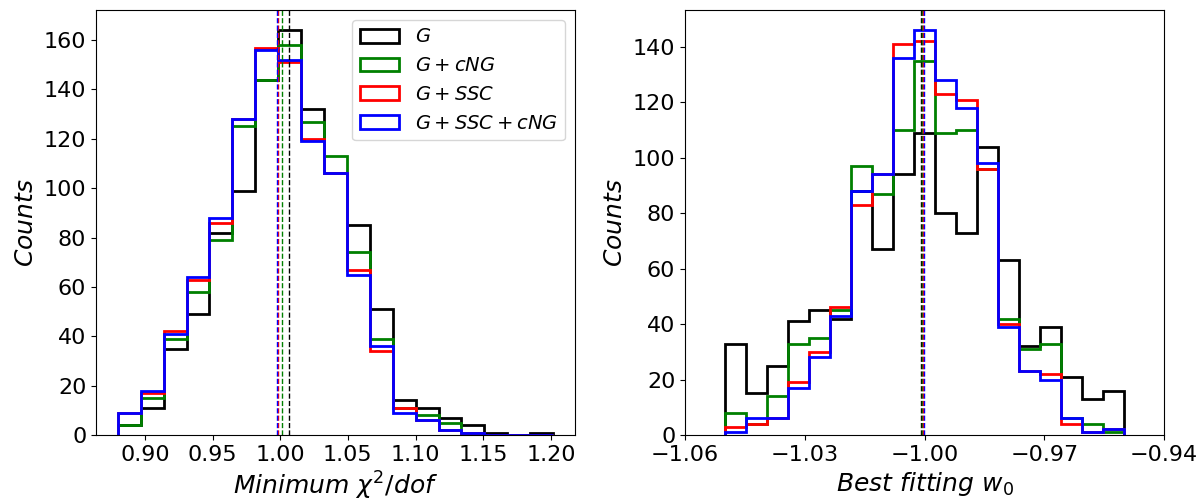}
        \caption{Distribution of the best-fitting $\chi^2$ (left; dof = $1100-1 = 1099$) and best-fitting $w_0$ (right) values from the analysis of $1000$ data vectors drawn from a multivariate Gaussian with the total G+SSC+cNG covariance. These $\chi^2$ values are evaluated as $\chi^2 = \left({\bf M} - {\bf D}\right) \cov^{-1} \left({\bf M} - {\bf D}\right)^{-1}$, where ${\bf D}$ is the data vector of our main Euclid-like setup ($s01$ in Table \ref{table:setups}) and ${\bf M}$ and $\cov$ are the corresponding model prediction and covariance matrix, respectively. Here, $w_0$ is the only varied parameter with the others held fixed at their fiducial values. The result is shown for constraints obtained with the G, G+cNG, G+SSC and G+SSC+cNG covariance subsets, as labeled. This test reveals the effect of a likelihood analysis that uses a different covariance than the one the data is drawn from. The vertical dashed lines indicate the mean values of the distributions (same color code).}
\label{fig:hists}
\end{figure}

Figure \ref{fig:s01} illustrates the relative contribution of the SSC and cNG terms to the total G+SSC+cNG covariance, for our main $N_{\rm tomo} = 10$ setup ($s01$ in Table \ref{table:setups}). The upper left and right panels show, respectively, the element-by-element ratio of the SSC and cNG contributions to the total result, as labeled. The lower panel shows the distribution of the matrix elements shown in the upper panels. Averaging over all the $1100^2$ elements of the covariance matrix, the cNG contribution amounts to $\approx 20\%$ of the total covariance matrix; the cNG term only contributes with $\approx 50\%$ of the total in a very small number of elements. This figure thus already indicates that the cNG term represents a subdominant contribution to the total covariance matrix. The relevant way to quantify the importance of the cNG term is however not at the level of the matrix elements, but instead at the level of final parameter constraints, which we turn to next\footnote{The interested reader can find in Ref.~\cite{completessc} a more detailed illustration of the relative contribution and scale dependence of the G, SSC and cNG terms for a single source redshift plane at $z_S = 1$.}.

Figure \ref{fig:n01_1d} shows the constraints obtained on each cosmological parameter when all the others are held fixed at their fiducial values. The result is shown for four covariance subsets: G, G+cNG, G+SSC and G+SSC+cNG, as labeled. The lower right panel summarizes the impact of the different covariance subsets on the unmarginalized $1\sigma$ confidence intervals for the five parameters. As expected, of the four subsets shown, the one that contains only the G term is that which yields the tightest constraints on parameters. Further, adding the cNG term to the G term results in an appreciable increase in the size of the error bars ($\approx 35\%$). Taking this observation at face value, one might conclude that the cNG must indeed be evaluated with high accuracy given its importance in setting the size of the error on parameter constraints. This would however be a premature conclusion since it ignores the contribution of the considerably more important SSC term. Indeed we see that adding the cNG term to the G+SSC subset only increases the size of the error bars by $\lesssim 5\%$.

Figure \ref{fig:n01_2d} shows the marginalized two-dimensional $2\sigma$ confidence contours on cosmological parameters when all parameters are varied in the constraints. The take-away message from Fig.~\ref{fig:n01_2d} is the same as that from Fig.~\ref{fig:n01_1d}: the SSC dominates the degradation of the constraints relative to the G component, or in other words, the addition of the cNG to the G+SSC covariance matrix results in negligible changes in parameter contours. The impact of the cNG term on the corresponding one-dimensional marginalized constraints (not shown) is even smaller than that shown in Fig.~\ref{fig:n01_1d} because the increase in the parameter ellipsoid volumes tends to take place proportionally to the degeneracy directions.\footnote{For completeness, we note that the Figure of Merit (${\rm FoM}$) of the full five-dimensional parameter space (defined as ${\rm FoM} = {\rm det}\left({\rm Cov}_{\rm params}\right)^{-1/2}$, where ${\rm Cov}_{\rm params}$ is the covariance matrix of the cosmological parameters) increases by $\approx 14\%$ if one drops the cNG term, a factor of $\approx 2$ if one drops the SSC term and a factor of $\approx 4$ if one drops both the SSC and the cNG term, relative to the FoM obtained with the full G+cNG+SSC matrix.} We can thus regard the result depicted in Fig.~\ref{fig:n01_1d} as an upper bound on the impact that the cNG term will actually have in real analyses when several parameters (including nuisance systematic parameters that we do not consider here) are simultaneously varied.

The results of Figs.~\ref{fig:n01_1d} and \ref{fig:n01_2d} correspond to simulated analyses using a noise-free realization of the data vector, i.e., the data vector matches exactly the theoretical prediction (cf.~Eq.~(\ref{eq:prediction})) at our fiducial cosmology. As another check of the importance of the cNG term we analyze noisy realizations of the data vector. Specifically, we draw 1000 data vectors from a multivariate Gaussian distribution with the G+SSC+cNG covariance matrix and with mean given by the fiducial theoretical prediction. We analyze these noisy data vectors and obtain simulated constraints on $w_0$ (with the other parameters fixed) using varying covariance subsets in the likelihood analysis. This analysis is intended to mimic a real-life analysis in which we observe a data vector that is a realization of a ``true'' covariance matrix (here the total G+SSC+cNG covariance), but choose to perform likelihood analyses with a ``wrong'' covariance matrix. The result is summarized in the histograms of Fig.~\ref{fig:hists}, which reveal no clear indication that the use of the wrong covariance matrix introduces a bias in the resulting goodness-of-fit ($\chi^2$ evaluated at the best-fit cosmology; left panel, $dof = 1099$) or best-fitting parameter value (right panel).

It is interesting to note that the same conclusion regarding the goodness of fit holds also for the G and G+cNG covariance matrices, despite Fig.~\ref{fig:n01_1d} showing that these covariance subsets underestimate the statistical error on $w_0$ by $\approx 50\%$ and $\approx 30\%$, respectively. This reflects the fact that changes in the structure of the covariance matrix, while having a large impact on parameter errors, do not necessarily have a strong impact on the location of the maximum of the posterior and on the overall goodness-of-fit.

In summary, the results discussed in this subsection tell us that \emph{owing to the dominance of the G and SSC terms, the cNG term contributes only a marginal amount to the parameter error bars in lensing tomography constraints, and neglecting it yields no visible changes to the resulting best-fitting values and goodness-of-fit.} These results are in line with the findings of Ref.~\cite{2009MNRAS.395.2065T} who used a Fisher matrix analysis to study the impact on parameter constraints of the cNG term and of an approximation to the SSC term (called ``non-linear beat coupling''); see also Refs.~\cite{2009ApJ...701..945S, completessc, 2018arXiv180511629T}, who presented further hints on the unimportance of the cNG term.

\subsection{The unimportance of the cNG term for varying survey parameters}\label{sec:res2}

\begin{figure}
        \centering
        \includegraphics[width=\textwidth]{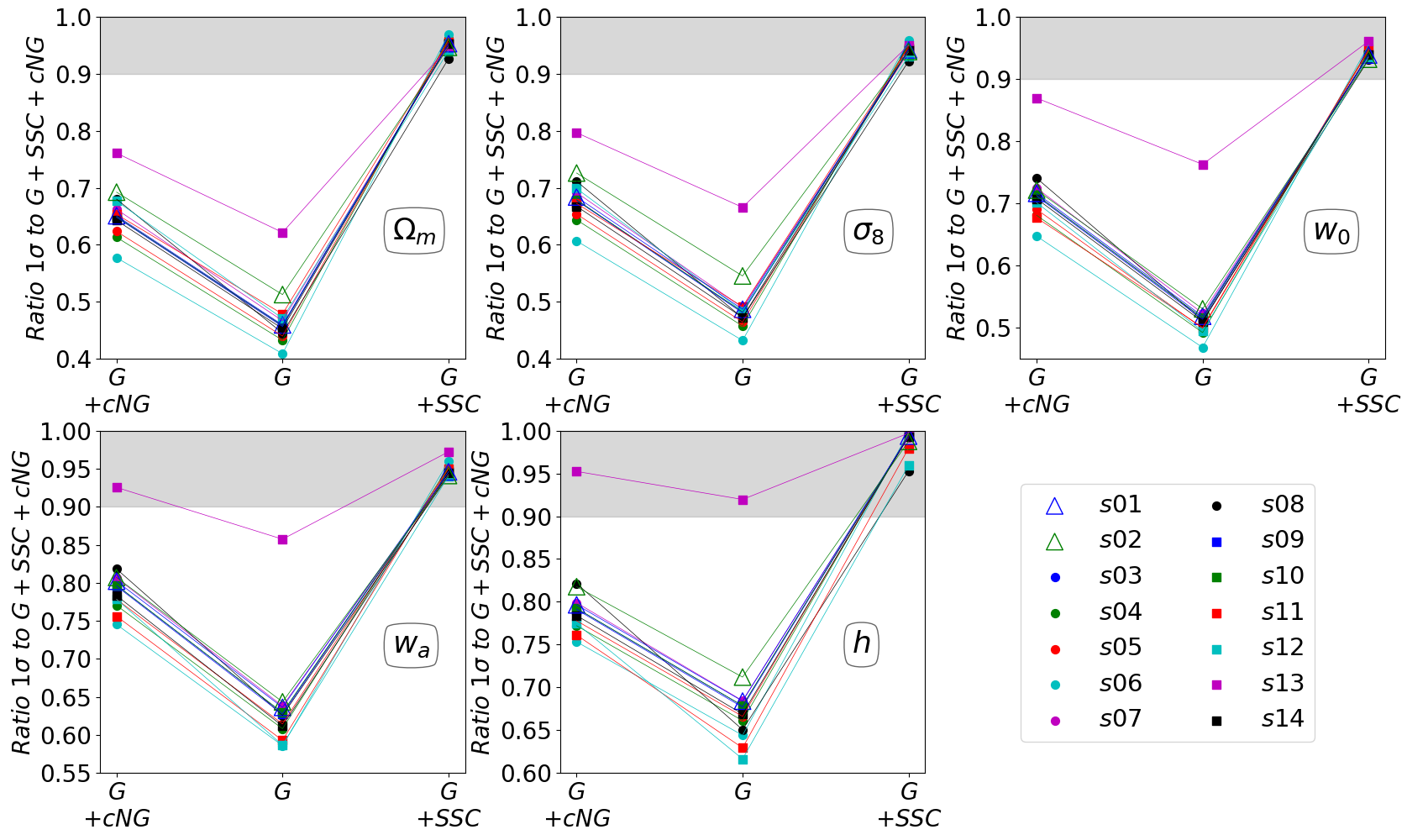}
        \caption{Ratio of the $1\sigma$ limits obtained with the G+cNG, G and G+SSC covariance subsets to that obtained with the total G+SSC+cNG covariance, for all the survey/analysis specifications listed in Table \ref{table:setups}, as labeled (the grey band marks $10\%$). We show ratios of unmarginalized parameter errors, with all other parameters held fixed at their fiducial values. The data vector used here corresponds to a noise-free realization matching the prediction at the fiducial cosmology. The various points are hard to distinguish, but the main point is that the G+SSC ones (right most points in each panel) all lie comfortably inside the $10\%$ band.
        }
\label{fig:setups_1d}
\end{figure}

In this subsection, we discuss how the conclusion drawn in the last subsection for our main Euclid-like setup holds for other survey/analysis specifications. Our findings are summarized in Fig.~\ref{fig:setups_1d}, which shows the ratios of the $1\sigma$ limits obtained for the G+cNG, G and G+SSC covariance subsets to that obtained using the total G+SSC+cNG covariance. With these variations we can test the impact of:

\begin{enumerate}

\item \underline{Euclid vs. LSST.} The case $s02$ corresponds to a case with $N_{\rm tomo} = 5$, but with the expected source galaxy distribution of the LSST survey, as well as its $f_{\rm sky}$ value.

\item \underline{Number of tomographic bins.} We tested three cases with $N_{\rm tomo} = 10, 5, 3$, which are labeled as $s01$, $s03$ and $s14$, respectively.

\item \underline{Mask shape.} For $N_{\rm tomo} = 5$, we ran constraints for three mask shapes: spherical polar cap, equatorial band and two spherical polar caps. These are cases $s03$, $s04$ and $s05$, respectively.

\item  \underline{Sky fraction, $f_{\rm sky}$.}  For $N_{\rm tomo} = 5$, the cases $s03$, $s06$ and $s07$ show constraints for $f_{\rm sky} = 0.36, 0.05, 0.50$, respectively.

\item \underline{Type of cNG calculation.} Comparing the results from the cases $s03$ and $s08$ shows the differences between the response and the halo model calculation of the cNG term.

\item  \underline{Number of $\ell$ bins.} For $N_{\rm tomo} = 5$, the cases $s03$, $s09$ and $s10$ show the differences between $N_{\ell} = 20, 10, 30$, respectively.

\item \underline{Range of scales.} We also tested the impact of varying the range of angular scales. In particular, relative to case $s03$, the minimum multipole considered in $s11$ increases from $\ell_{\rm min} = 20$ to $500$ and the maximum multipole from $\ell_{\rm max} = 5000$ to 10000 (case $s12$ further switches from the cNG response calculation to that of the halo model). This restriction to smaller angular scales reduces the relative importance of the G contribution. On the other hand, case $s13$ has the maximum multipole reduced from $\ell = 5000$ to $1000$, which enhances the relative contribution from the G term.

\end{enumerate}

Naturally, these different setups can have a marked impact on the size of the error bars themselves, but here we are interested in the relative differences for varying covariance subsets. The result depicted in Fig.~\ref{fig:setups_1d} shows that, for all parameters and for all of the setups tested, the cNG covariance always contributes a marginal amount to the total error on parameters: ignoring the cNG contribution still yields error bars that are within $8\%$ of those obtained with the total covariance matrix. This shows that the unimportance of the cNG term for our main Euclid-like setup discussed in the previous section holds also for a number of variations around it\footnote{It is interesting to note that case $s13$ stands out from the others in Fig.~\ref{fig:setups_1d}. This is expected since the cut of small angular scales that characterizes this survey setup (cf.~Table \ref{table:setups}) suppresses the contribution of the two non-Gaussian terms (cNG and SSC), and consequently, dropping each or both from the analysis has a smaller impact on the error. For instance, for case $s13$, the error bars on $h$ obtained with all covariance subsets are within $10\%$ of one another.}.

As an additional test, we have repeated the analysis of the setup $s03$, but using the real space correlation function as data vector (and associated covariance matrix) . In this case, we have further used $30$ angular bins with $\theta_{\rm max} = \pi/\ell_{\rm min}$, and $\theta_{\rm min} = \pi/\ell_{\rm max}$ for $\xi_+$ and $\theta_{\rm min} = 10 \pi/\ell_{\rm max}$ for $\xi_-$. For all five cosmological parameters, we have found that neglecting the cNG contribution decreases the error bars by less than $5\%$. This is in accordance with the power spectrum results, as expected.

\section{Summary and Discussion}\label{sec:conc}

We have examined the relative impact on parameter constraints of the three physical contributions to the covariance matrix of weak lensing two-point statistics: Gaussian (G), super-sample covariance (SSC) and connected non-Gaussian (cNG) terms (cf.~Sec.~\ref{sec:covdec}). More specifically, we focused on Euclid-like ($s01$ case in Table \ref{table:setups}) and LSST-like ($s02$) survey specifications, as well as variations around them, and have carried out forecast exercises for tomographic lensing convergence data vectors to analyze how various covariance subsets affect the resulting size of the parameter error bars. We have explored constraints on five cosmological parameters, $\{\Omega_{\rm m}, \sigma_8, w_0, w_a, h\}$, which we have separately and jointly constrained. 

Our main results can be summarized as follows:

\begin{itemize}

\item At the level of one-dimensional parameter constraints in our main Euclid-like setup, the error bars obtained with a G+SSC+cNG covariance matrix are only $\lesssim 5\%$ larger than those obtained with a G+SSC covariance matrix (cf.~Fig.~\ref{fig:n01_1d}). This demonstrates the relative unimportance of the cNG term in determining parameter error bars, which is also manifest when all cosmological parameters are constrained simultaneously (cf.~Fig.~\ref{fig:n01_2d}).

\item Dropping the cNG term from the constraints of several realizations of the data vector drawn from a multivariate Gaussian with the total G+SSC+cNG covariance did not reveal any bias at the level of the overall goodness-of-fit nor best-fitting parameter values (cf.~Fig.~\ref{fig:hists}).

\item The unimportance of the cNG term prevailed in our results even after exploring a number of variations in the analysis specifications including source redshift distribution, number of tomographic bins, mask shape, $f_{\rm sky}$, range of scales, number of multipole bins, configuration vs.~Fourier space, as well as different analytical recipes to the cNG term (response approach vs.~halo model). In all cases, removing the cNG contribution from the covariance resulted in a change in parameter errors of at most $8\%$.

\item In our forecasts, we did not consider the impact of systematic errors, which would increase the total error budget and hence further suppress the importance of the cNG term. Our conclusions on the unimportance of the cNG term can thus be regarded as conservative upper bounds.

\end{itemize}

\bigskip 

The observation that the cNG contribution is much less important than that of the G and SSC terms does not justify neglecting its contribution entirely, especially if it can be straightforwardly calculated using the response approach. However, the relative unimportance of the cNG term does relax the accuracy and precision requirements on its evaluation. As discussed in Sec.~\ref{sec:cNGterm}, a main concern about non-ensemble approaches to the cNG term is that they can at most provide approximations to the true cNG contribution. The key question to address, however, is: \emph{do these inadequacies matter at the level of parameter constraints?} The results presented in this paper suggest that the answer is no: the difference between using the response approach result or the ``true'' cNG contribution in constraints with the total G+SSC+cNG covariance matrix is likely to result in nearly indistinguishable parameter errors.

Another common worry about analytical approaches to the cNG term concerns the difficulties in incorporating more detailed survey specifications such as non-contiguous masks (i.e., masks with holes) or varying survey depth (i.e. $n_{S}(z)$ becomes also a function of $\vtheta$). Indeed, the inclusion of these effects is not as straightforward as the calculation presented here, but it is possible to conceive ways to take them into account. We note, however, that the inclusion of these effects would represent corrections to a term that is responsible for an already small ($\lesssim 5\%$) contribution to parameter errors; these corrections would thus likely be negligible. Investigating the size of these effects in the G and SSC terms remains an interesting important task to carry out, but these terms are not as numerically demanding as the cNG term. For instance, ensembles of Gaussian realizations can be generated inexpensively to estimate the G term in such nontrivial survey specifications.\footnote{For completeness, nontrivial survey mask geometries can also have an important impact on the shape noise contribution, but similarly, this can be efficiently addressed \cite{2011ApJ...734...76S, 2018arXiv180410663T}.} It is also possible to conceive of ways to generalize the SSC formulae of Sec.~\ref{sec:SSCterm} to include these effects.

Another interesting issue in parameter inference using weak lensing data that has recently become of interest concerns the shape of the likelihood function. In this paper, we have assumed it to be a multivariate Gaussian, but the degree to which this is a valid approximation should of course be put to test \cite{2016MNRAS.456L.132S, 2018MNRAS.473.2355S, 2013A&A...551A..88C}. Recently, Ref.~\cite{2018MNRAS.477.4879S} have taken a few steps in this direction and found that the Gaussian likelihood approximation can indeed break down, especially on large scales due to the small number of degrees of freedom involved in the two-point function (which invalidates the Gaussian approximation). The full, correct shape of the likelihood should nonetheless still be described by a generalized covariance matrix, for which our conclusions on the size of the cNG term should equally apply. We note also that the setups $s11$ and $s12$ in Table \ref{table:setups} drop the largest angular scales and are thus less susceptible to be affected by a breakdown of the Gaussian likelihood assumption.

In this paper, we devoted our attention to the two-point function of weak lensing observables, but two-point statistics of galaxy distributions (either spectroscopic or photometric), and corresponding cross-correlations with lensing, are further important data vectors that future wide field imaging surveys will measure. In these cases, the covariance matrix can likewise be decomposed into the G, SSC and cNG terms, and hence, it is desirable to also learn about their relative importance in order to pinpoint which contributions are worthy of more or less attention in galaxy covariance estimation (similar lines of reasoning apply to the case of higher-order $N$-point functions as well). A few interesting steps in this direction were taken recently by Refs.~\cite{2018arXiv180609497B, 2018arXiv180609477L} who, taking the redshift-space dark matter halo power spectrum and correlation function as data vectors (see also Ref.~\cite{2018arXiv180609499C} for the halo bispectrum), compared the parameter constraints obtained using G+cNG subsets evaluated with various methods, including one that considers only the G term.

\begin{acknowledgments}

We thank Benjamin Joachimi, Fabien Lacasa and Emmanuel Schaan for very useful comments and discussions. 
Some of the results in this paper were obtained using the {\sc Healpix} package \cite{2005ApJ...622..759G}.
EK acknowledges support from NASA grant 15-WFIRST15-0008 Cosmology with the High Latitude Survey WFIRST Science Investigation Team. Part of the research was carried out at the Jet Propulsion Laboratory, California Institute of Technology, under a contract with the National Aeronautics and Space Administration.
FS acknowledges support from the Starting Grant (ERC-2015-STG 678652) ``GrInflaGal'' of the European Research Council.
\end{acknowledgments}

\bibliography{REFS}

\def\eprinttmppp@#1arXiv:@{#1}
\providecommand{\arxivlink[1]}{\href{http://arxiv.org/abs/#1}{arXiv:#1}}
\providecommand{\arxivlinknopre[1]}{\href{http://arxiv.org/abs/#1}{#1}}
\providecommand{\eprintmod}[1][XXXX.XXXX]{\IfSubStr{#1}{arXiv}{\arxivlinknopre{#1}}{\arxivlink{#1}}}
\providecommand{\adsurl}[1]{\href{#1}{ADS}}
\begin{thebibliography}{82}
\expandafter\ifx\csname natexlab\endcsname\relax\def\natexlab#1{#1}\fi
\expandafter\ifx\csname bibnamefont\endcsname\relax
  \def\bibnamefont#1{#1}\fi
\expandafter\ifx\csname bibfnamefont\endcsname\relax
  \def\bibfnamefont#1{#1}\fi
\expandafter\ifx\csname citenamefont\endcsname\relax
  \def\citenamefont#1{#1}\fi
\expandafter\ifx\csname url\endcsname\relax
  \def\url#1{\texttt{#1}}\fi
\expandafter\ifx\csname urlprefix\endcsname\relax\def\urlprefix{URL }\fi

\bibitem{2017MNRAS.465.1454H}
H.~{Hildebrandt} {\em et~al.},
\newblock MNRAS {\bf 465}, 1454 (2017), [\eprintmod[1606.05338]].

\bibitem{2017arXiv170706627J}
S.~{Joudaki} {\em et~al.},
\newblock ArXiv e-prints  (2017), [\eprintmod[1707.06627]].

\bibitem{2017arXiv170605004V}
E.~{van Uitert} {\em et~al.},
\newblock ArXiv e-prints  (2017), [\eprintmod[1706.05004]].

\bibitem{diehl/etal:2014}
H.~T. {Diehl} {\em et~al.},
\newblock {The Dark Energy Survey and operations: Year 1},
\newblock in {\em Observatory Operations: Strategies, Processes, and Systems
  V}, , \procspie Vol. 9149, p. 91490V, 2014.

\bibitem{2017arXiv170801530D}
{DES Collaboration} {\em et~al.},
\newblock ArXiv e-prints  (2017), [\eprintmod[1708.01530]].

\bibitem{2018PASJ...70S..25M}
R.~{Mandelbaum} {\em et~al.},
\newblock PASJ {\bf 70}, S25 (2018), [\eprintmod[1705.06745]].

\bibitem{2011arXiv1110.3193L}
R.~{Laureijs} {\em et~al.},
\newblock ArXiv e-prints  (2011), [\eprintmod[1110.3193]].

\bibitem{2012arXiv1211.0310L}
{LSST Dark Energy Science Collaboration},
\newblock ArXiv e-prints  (2012), [\eprintmod[1211.0310]].

\bibitem{2013arXiv1305.5422S}
D.~{Spergel} {\em et~al.},
\newblock ArXiv e-prints  (2013), [\eprintmod[1305.5422]].

\bibitem{takada/hu:2013}
M.~Takada and W.~Hu,
\newblock Phys.Rev. {\bf D87}, 123504 (2013), [\eprintmod[1302.6994]].

\bibitem{li/hu/takada}
Y.~{Li}, W.~{Hu} and M.~{Takada},
\newblock \prd {\bf 89}, 083519 (2014), [\eprintmod[1401.0385]].

\bibitem{2014PhRvD..90j3530L}
Y.~{Li}, W.~{Hu} and M.~{Takada},
\newblock \prd {\bf 90}, 103530 (2014), [\eprintmod[1408.1081]].

\bibitem{2014PhRvD..90b3003M}
A.~{Manzotti}, W.~{Hu} and A.~{Benoit-L{\'e}vy},
\newblock PRD {\bf 90}, 023003 (2014), [\eprintmod[1401.7992]].

\bibitem{2014MNRAS.441.2456T}
M.~{Takada} and D.~N. {Spergel},
\newblock MNRAS {\bf 441}, 2456 (2014), [\eprintmod[1307.4399]].

\bibitem{2014MNRAS.444.3473T}
R.~{Takahashi}, S.~{Soma}, M.~{Takada} and I.~{Kayo},
\newblock MNRAS {\bf 444}, 3473 (2014), [\eprintmod[1405.2666]].

\bibitem{1999ApJ...527....1S}
R.~{Scoccimarro}, M.~{Zaldarriaga} and L.~{Hui},
\newblock APJ {\bf 527}, 1 (1999), [\eprintmod[astro-ph/9901099]].

\bibitem{2001ApJ...554...56C}
A.~{Cooray} and W.~{Hu},
\newblock \apj {\bf 554}, 56 (2001), [\eprintmod[astro-ph/0012087]].

\bibitem{2009MNRAS.395.2065T}
M.~{Takada} and B.~{Jain},
\newblock \mnras {\bf 395}, 2065 (2009), [\eprintmod[0810.4170]].

\bibitem{2016JCAP...06..052B}
D.~{Bertolini}, K.~{Schutz}, M.~P. {Solon} and K.~M. {Zurek},
\newblock \jcap {\bf 6}, 052 (2016), [\eprintmod[1604.01770]].

\bibitem{2016PhRvD..93l3505B}
D.~{Bertolini}, K.~{Schutz}, M.~P. {Solon}, J.~R. {Walsh} and K.~M. {Zurek},
\newblock \prd {\bf 93}, 123505 (2016), [\eprintmod[1512.07630]].

\bibitem{mohammed1}
I.~{Mohammed}, U.~{Seljak} and Z.~{Vlah},
\newblock \mnras {\bf 466}, 780 (2017), [\eprintmod[1607.00043]].

\bibitem{responses1}
A.~{Barreira} and F.~{Schmidt},
\newblock JCAP {\bf 6}, 053 (2017), [\eprintmod[1703.09212]].

\bibitem{responses2}
A.~{Barreira} and F.~{Schmidt},
\newblock \jcap {\bf 11}, 051 (2017), [\eprintmod[1705.01092]].

\bibitem{emulator}
K.~{Heitmann}, E.~{Lawrence}, J.~{Kwan}, S.~{Habib} and D.~{Higdon},
\newblock \apj {\bf 780}, 111 (2014), [\eprintmod[1304.7849]].

\bibitem{completessc}
A.~{Barreira}, E.~{Krause} and F.~{Schmidt},
\newblock \jcap {\bf 6}, 015 (2018), [\eprintmod[1711.07467]].

\bibitem{wagner/etal:2014}
C.~Wagner, F.~Schmidt, C.-T. Chiang and E.~Komatsu,
\newblock Mon.Not.Roy.Astron.Soc. {\bf 448}, 11 (2015),
  [\eprintmod[1409.6294]].

\bibitem{CFCpaper2}
L.~{Dai}, E.~{Pajer} and F.~{Schmidt},
\newblock \jcap {\bf 10}, 059 (2015), [\eprintmod[1504.00351]].

\bibitem{li/hu/takada:2016}
Y.~{Li}, W.~{Hu} and M.~{Takada},
\newblock \prd {\bf 93}, 063507 (2016), [\eprintmod[1511.01454]].

\bibitem{lazeyras/etal}
T.~{Lazeyras}, C.~{Wagner}, T.~{Baldauf} and F.~{Schmidt},
\newblock \jcap {\bf 2}, 018 (2016), [\eprintmod[1511.01096]].

\bibitem{response}
C.~{Wagner}, F.~{Schmidt}, C.-T. {Chiang} and E.~{Komatsu},
\newblock \jcap {\bf 8}, 042 (2015), [\eprintmod[1503.03487]].

\bibitem{andreas}
A.~S. {Schmidt}, S.~D.~M. {White}, F.~{Schmidt} and J.~{St{\"u}cker},
\newblock ArXiv e-prints  (2018), [\eprintmod[1803.03274]].

\bibitem{bertolini1}
D.~{Bertolini} and M.~P. {Solon},
\newblock JCAP {\bf 11}, 030 (2016), [\eprintmod[1608.01310]].

\bibitem{cooray/sheth}
A.~Cooray and R.~K. Sheth,
\newblock Phys.Rept. {\bf 372}, 1 (2002), [\eprintmod[astro-ph/0206508]].

\bibitem{cooray/hu:2001_cov}
A.~{Cooray} and W.~{Hu},
\newblock \apj {\bf 554}, 56 (2001), [\eprintmod[astro-ph/0012087]].

\bibitem{2016PhRvD..93f3512S}
F.~{Schmidt},
\newblock \prd {\bf 93}, 063512 (2016), [\eprintmod[1511.02231]].

\bibitem{2009ApJ...700..479T}
R.~{Takahashi} {\em et~al.},
\newblock \apj {\bf 700}, 479 (2009), [\eprintmod[0902.0371]].

\bibitem{2009ApJ...701..945S}
M.~{Sato} {\em et~al.},
\newblock \apj {\bf 701}, 945 (2009), [\eprintmod[0906.2237]].

\bibitem{2011ApJ...734...76S}
M.~{Sato}, M.~{Takada}, T.~{Hamana} and T.~{Matsubara},
\newblock \apj {\bf 734}, 76 (2011), [\eprintmod[1009.2558]].

\bibitem{2012MNRAS.426.1262H}
J.~{Harnois-D{\'e}raps}, S.~{Vafaei} and L.~{Van Waerbeke},
\newblock MNRAS {\bf 426}, 1262 (2012), [\eprintmod[1202.2332]].

\bibitem{blot2015}
L.~{Blot}, P.~S. {Corasaniti}, J.-M. {Alimi}, V.~{Reverdy} and Y.~{Rasera},
\newblock MNRAS {\bf 446}, 1756 (2015), [\eprintmod[1406.2713]].

\bibitem{2017ApJ...850...24T}
R.~{Takahashi} {\em et~al.},
\newblock APJ {\bf 850}, 24 (2017), [\eprintmod[1706.01472]].

\bibitem{2016PhRvD..93f3524P}
A.~{Petri}, Z.~{Haiman} and M.~{May},
\newblock \prd {\bf 93}, 063524 (2016), [\eprintmod[1601.06792]].

\bibitem{2018MNRAS.478.4602K}
A.~{Klypin} and F.~{Prada},
\newblock \mnras {\bf 478}, 4602 (2018), [\eprintmod[1701.05690]].

\bibitem{2018arXiv180504511H}
J.~{Harnois-Deraps} {\em et~al.},
\newblock ArXiv e-prints  (2018), [\eprintmod[1805.04511]].

\bibitem{2016MNRAS.459.2327I}
A.~{Izard}, M.~{Crocce} and P.~{Fosalba},
\newblock \mnras {\bf 459}, 2327 (2016), [\eprintmod[1509.04685]].

\bibitem{2018arXiv180105745S}
R.~{Sgier}, A.~{R{\'e}fr{\'e}gier}, A.~{Amara} and A.~{Nicola},
\newblock ArXiv e-prints  (2018), [\eprintmod[1801.05745]].

\bibitem{2017JCAP...01..008R}
L.~A. {Rizzo} {\em et~al.},
\newblock \jcap {\bf 1}, 008 (2017), [\eprintmod[1610.07624]].

\bibitem{2007A&A...464..399H}
J.~{Hartlap}, P.~{Simon} and P.~{Schneider},
\newblock \aap {\bf 464}, 399 (2007), [\eprintmod[astro-ph/0608064]].

\bibitem{2013MNRAS.432.1928T}
A.~{Taylor}, B.~{Joachimi} and T.~{Kitching},
\newblock \mnras {\bf 432}, 1928 (2013), [\eprintmod[1212.4359]].

\bibitem{2001PhR...340..291B}
M.~{Bartelmann} and P.~{Schneider},
\newblock PhysRep {\bf 340}, 291 (2001), [\eprintmod[astro-ph/9912508]].

\bibitem{2005astro.ph..9252S}
P.~{Schneider},
\newblock ArXiv Astrophysics e-prints  (2005), [\eprintmod[astro-ph/0509252]].

\bibitem{2008ARNPS..58...99H}
H.~{Hoekstra} and B.~{Jain},
\newblock Annual Review of Nuclear and Particle Science {\bf 58}, 99 (2008),
  [\eprintmod[0805.0139]].

\bibitem{2015RPPh...78h6901K}
M.~{Kilbinger},
\newblock Reports on Progress in Physics {\bf 78}, 086901 (2015),
  [\eprintmod[1411.0115]].

\bibitem{2003MNRAS.341.1311S}
R.~E. {Smith} {\em et~al.},
\newblock \mnras {\bf 341}, 1311 (2003), [\eprintmod[astro-ph/0207664]].

\bibitem{2012ApJ...761..152T}
R.~{Takahashi}, M.~{Sato}, T.~{Nishimichi}, A.~{Taruya} and M.~{Oguri},
\newblock \apj {\bf 761}, 152 (2012), [\eprintmod[1208.2701]].

\bibitem{2013LRR....16....6A}
L.~{Amendola} {\em et~al.},
\newblock Living Reviews in Relativity {\bf 16}, 6 (2013),
  [\eprintmod[1206.1225]].

\bibitem{chang/etal:2013}
C.~{Chang} {\em et~al.},
\newblock \mnras {\bf 434}, 2121 (2013), [\eprintmod[1305.0793]].

\bibitem{DESC-SRD}
{The LSST Dark Energy Science Collaboration} {\em et~al.},
\newblock ArXiv e-prints  (2018), [\eprintmod[1809.01669]].

\bibitem{2018arXiv180511629T}
R.~{Takahashi}, T.~{Nishimichi}, M.~{Takada}, M.~{Shirasaki} and
  K.~{Shiroyama},
\newblock ArXiv e-prints  (2018), [\eprintmod[1805.11629]].

\bibitem{2018arXiv180410663T}
M.~A. {Troxel} {\em et~al.},
\newblock ArXiv e-prints  (2018), [\eprintmod[1804.10663]].

\bibitem{2004MNRAS.349..603E}
G.~{Efstathiou},
\newblock \mnras {\bf 349}, 603 (2004), [\eprintmod[astro-ph/0307515]].

\bibitem{2002ApJ...567....2H}
E.~{Hivon} {\em et~al.},
\newblock \apj {\bf 567}, 2 (2002), [\eprintmod[astro-ph/0105302]].

\bibitem{2017arXiv170609359K}
E.~{Krause} {\em et~al.},
\newblock ArXiv e-prints  (2017), [\eprintmod[1706.09359]].

\bibitem{2017MNRAS.470.2100K}
E.~{Krause} and T.~{Eifler},
\newblock MNRAS {\bf 470}, 2100 (2017), [\eprintmod[1601.05779]].

\bibitem{2006MNRAS.371.1188H}
A.~J.~S. {Hamilton}, C.~D. {Rimes} and R.~{Scoccimarro},
\newblock MNRAS {\bf 371}, 1188 (2006), [\eprintmod[astro-ph/0511416]].

\bibitem{2006PhRvD..74b3522S}
E.~{Sefusatti}, M.~{Crocce}, S.~{Pueblas} and R.~{Scoccimarro},
\newblock \prd {\bf 74}, 023522 (2006), [\eprintmod[arXiv:astro-ph/0604505]].

\bibitem{2012JCAP...04..019D}
R.~{de Putter}, C.~{Wagner}, O.~{Mena}, L.~{Verde} and W.~J. {Percival},
\newblock JCAP {\bf 4}, 019 (2012), [\eprintmod[1111.6596]].

\bibitem{2007NJPh....9..446T}
M.~{Takada} and S.~{Bridle},
\newblock New Journal of Physics {\bf 9}, 446 (2007), [\eprintmod[0705.0163]].

\bibitem{2013MNRAS.429..344K}
I.~{Kayo}, M.~{Takada} and B.~{Jain},
\newblock \mnras {\bf 429}, 344 (2013), [\eprintmod[1207.6322]].

\bibitem{2003ApJ...584..702H}
W.~{Hu} and A.~V. {Kravtsov},
\newblock \apj {\bf 584}, 702 (2003), [\eprintmod[astro-ph/0203169]].

\bibitem{2016arXiv161205958L}
F.~{Lacasa}, M.~{Lima} and M.~{Aguena},
\newblock ArXiv e-prints  (2016), [\eprintmod[1612.05958]].

\bibitem{2018JCAP...02..022L}
Y.~{Li}, M.~{Schmittfull} and U.~{Seljak},
\newblock \jcap {\bf 2}, 022 (2018), [\eprintmod[1711.00018]].

\bibitem{2009MNRAS.398.1601F}
F.~{Feroz}, M.~P. {Hobson} and M.~{Bridges},
\newblock \mnras {\bf 398}, 1601 (2009), [\eprintmod[0809.3437]].

\bibitem{2016MNRAS.456L.132S}
E.~{Sellentin} and A.~F. {Heavens},
\newblock MNRAS {\bf 456}, L132 (2016), [\eprintmod[1511.05969]].

\bibitem{2018MNRAS.473.2355S}
E.~{Sellentin} and A.~F. {Heavens},
\newblock \mnras {\bf 473}, 2355 (2018), [\eprintmod[1707.04488]].

\bibitem{2013A&A...551A..88C}
J.~{Carron},
\newblock \aap {\bf 551}, A88 (2013), [\eprintmod[1204.4724]].

\bibitem{2018MNRAS.477.4879S}
E.~{Sellentin}, C.~{Heymans} and J.~{Harnois-D{\'e}raps},
\newblock \mnras {\bf 477}, 4879 (2018), [\eprintmod[1712.04923]].

\bibitem{2018arXiv180609497B}
L.~{Blot} {\em et~al.},
\newblock ArXiv e-prints  (2018), [\eprintmod[1806.09497]].

\bibitem{2018arXiv180609477L}
M.~{Lippich} {\em et~al.},
\newblock ArXiv e-prints  (2018), [\eprintmod[1806.09477]].

\bibitem{2018arXiv180609499C}
M.~{Colavincenzo} {\em et~al.},
\newblock ArXiv e-prints  (2018), [\eprintmod[1806.09499]].

\bibitem{2005ApJ...622..759G}
K.~M. {G{\'o}rski} {\em et~al.},
\newblock \apj {\bf 622}, 759 (2005), [\eprintmod[astro-ph/0409513]].

\end{thebibliography}

\end{document}